\documentclass[draft]{IEEEtran}
\usepackage{cite}
\usepackage{amsmath,amssymb,amsfonts}
\usepackage{algorithmic}
\usepackage{graphicx}
\usepackage{textcomp}
\usepackage[caption=false]{subfig}
\def\BibTeX{{\rm B\kern-.05em{\sc i\kern-.025em b}\kern-.08em T\kern-.1667em\lower.7ex\hbox{E}\kern-.125emX}}

\begin{document}
\title{Steady-state Simulation of Semiconductor Devices using Discontinuous Galerkin Methods}
\author{Liang Chen and Hakan Bagci
\thanks{The authors are with the Division of Computer, Electrical, and Mathematical Sciences and Engineering, King Abdullah University of Science and Technology (KAUST), Thuwal 23955-6900, Saudi Arabia (e-mails:\{liang.chen, hakan.bagci\}@kaust.edu.sa).}}
\maketitle

\begin{abstract}
Design of modern nanostructured semiconductor devices often calls for simulation tools capable of modeling arbitrarily-shaped multiscale geometries. In this work, to this end, a discontinuous Galerkin (DG) method-based framework is developed to simulate steady-state response of semiconductor devices. The proposed framework solves a system of Poisson equation (in electric potential) and drift-diffusion equations (in charge densities), which are nonlinearly coupled via the drift current and the charge distribution. This system is ``decoupled’’ and ``linearized’’ using the Gummel method and the resulting equations are discretized using a local DG scheme. The proposed framework is used to simulate geometrically intricate semiconductor devices with realistic models of mobility and recombination rate. Its accuracy is demonstrated by comparing the results to those obtained by the finite volume and finite element methods implemented in a commercial software package.
\end{abstract}


\section{Introduction}
Simulation tools capable of numerically characterizing semiconductor devices play a vital role in device/system design frameworks used by the electronics industry as well as various related research fields~\cite{Selberher1984, Vasileska2010, Fu2014, Szymanski2017, Li2017, Wang2019, Nagy2019, Rossi2019}. Indeed, in the last several decades, numerous commercial and open source technology computer aided design (TCAD) tools, which implement various transport models ranging from semi-classical to quantum mechanical models, have been developed for this purpose~\cite{tcad}. Despite the recent trend of device miniaturization that requires simulators to account for quantum transport effects, many devices with larger dimensions (at the scale of $1\mu {\rm m}$) and with more complex geometries are being designed and implemented for various applications. Examples of these nanostructured devices range from photodiodes and phototransistors to solar cells, light emitting diodes, and photoconductive antennas~\cite{Piprek2018}. Electric field-charge carrier interactions on these devices can still be accurately accounted for using semi-classical models, however, their numerical simulation in TCAD raises challenges due to the presence of multi-scale and intricate geometric features.

Among the semi-classical approaches developed for modeling charge carrier transport, the drift-diffusion (DD) model is among the most popular ones because of its simplicity while being capable of explaining many essential characteristics of semiconductor devices~\cite{Selberher1984, Vasileska2010, Fu2014}. One well-known challenge in using the DD model is the exponential variation of carrier densities, which renders standard numerical schemes used for discretizing the model unstable unless an extremely fine mesh is used. This challenge traces back to the convection-dominated convection-diffusion equations, whose solutions show sharp boundary layers. Various stabilization techniques have been proposed and incorporated with different discretization schemes~\cite{Scharfetter1969, Bank1983, Sacco1997, Bank1998, Hillairet2003, Chatard2012, Chatard2012_1, Brezzi1989, Xu1999, Lazarov2005, Mauri2015}. The Scharfetter-Gummel (SG) method~\cite{Scharfetter1969} has been one of the workhorses in semiconductor device modeling; it  uses exponential functions to approximate the carrier densities so that the fine mesh requirement can be alleviated. The SG method has been first proposed for finite difference discretization, and then generalized to finite volume method (FVM)~\cite{Bank1983, Sacco1997, Bank1998, Hillairet2003, Chatard2012, Chatard2012_1} and finite element method (FEM)~\cite{Brezzi1989, Xu1999, Lazarov2005, Mauri2015}.

As mentioned above, many modern devices involve geometrically intricate structures. Therefore, FVM and FEM, which allow for unstructured meshes, have drawn more attention in recent years. However, the SG generalizations making use of FVM and FEM pose requirements on the regularity of the mesh~\cite{Bank1998, Chatard2012, Lazarov2005, Mauri2015, Farrell2018}. For example, FVM requires boundary conforming Delaunay triangulations for two dimensional (2D) simulations and admissible partitions for three dimensional (3D) ones~\cite{Bank1998, Chatard2012, Farrell2018}. These requirements cannot be easily satisfied in mesh generation for devices with complex geometries~\cite{Mauri2015, Farrell2018}. In addition, FEM stabilization techniques, such as the streamline upwind Petrov-Galerkin (SUPG) method~\cite{Hughes1979, Brooks1982} and the Galerkin least-square (GLS) method~\cite{Hughes1989, Franca1991}, have been used in simulation of semiconductor devices. However, SUPG suffers from ``artificial’’ numerical diffusion~\cite{Carey2000, Stynes2005, Ngo2015}; and GLS leads to unphysical smearing of the boundary layers and does not preserve current conservation~\cite{Carey2000, COMSOL}.

Although significant effort has been put into the numerical solution of the convection-dominated convection-diffusion problem in the last three decades, especially in the applied mathematics community, a fully-satisfactory numerical scheme for general industrial problems is yet to be formulated and implemented, for example see~\cite{Carey2000, Stynes2005, Brezzi2005, Roos2008, Roos2012} for surveys.

Recently, the discontinuous Galerkin (DG) method has attracted significant attention in several fields of computational science~\cite{Cockburn1999, Beatrice2008, Hesthaven2008, Pietro2012, Shu2016}. DG can be thought of as a hybrid method that combines the advantages of FVM and FEM. It uses local high-order expansions to represent/approximate the unknowns to be solved for. Each of these expansions is defined on a single mesh element and is ``connected’’ to other expansions defined on the neighboring elements via numerical flux. This approach equips DG with several advantages: The order of the local expansions can be changed individually, the mesh can be non-conformal (in addition to being unstructured), and the numerical flux can be designed to control the stability and accuracy characteristics of the DG scheme. More specifically, for semiconductor device simulations, the instability caused by the boundary layers can be alleviated without introducing much numerical diffusion. We should note here that for a given order of expansion $p$, DG requires a larger number of unknowns than FEM. However, the difference decreases as $p$ gets larger, and for many problems, DG benefits from  h- and/or p-refinement schemes~\cite{Hesthaven2008, Shu2016} and easily compensate for the small increase in the computational cost.

Those properties render DG an attractive option for multi-scale simulations~\cite{Cockburn1999, Beatrice2008, Hesthaven2008, Pietro2012, Shu2016, Augustin2011, Ngo2015}, and indeed, time domain DG has been recently used for transient semiconductor simulations~\cite{Liu2004, Liu2007, Liu2016}. However, in device TCAD, the non-equilibrium steady-state response of semiconductor devices is usually the most concerned case and it is computationally very costly to model in time domain because the simulation has to be executed for a very large number of time steps to reach the steady-state~\cite{Markowich1986, Vasileska2010}.

The steady-state simulation calls for solution of a nonlinear system consisting of three coupled second-order elliptic partial differential equations (PDEs). The first of these equations is the Poisson equation in scalar potential and the other two are the convection-diffusion type DD equations in electron and hole densities. These three equations are nonlinearly coupled via the drift current and the charge distribution. The charge-density dependent recombination rate, together with the field-dependent mobility and diffusion coefficients, makes the nonlinearity even stronger.  In this work, for the first time, a DG-based numerical framework is formulated and implemented to solve this coupled nonlinear system of equations. More specifically, we use the local DG (LDG) scheme~\cite{Cockburn1998} in cooperation with the Gummel method~\cite{Gummel1964} to simulate the non-equilibrium steady-state response of semiconductor devices.  To construct the (discretized) DG operator for the  convection-diffusion type DD equations (linearized within the Gummel method), the LDG alternate numerical flux is used for the diffusion term~\cite{Castillo2001} and the local Lax-Friedrichs flux is used for the convection term. Similarly, the discretized DG operator for the Poisson equation (linearized within the Gummel method) is constructed using the alternate numerical flux. The resulting DG-based framework is used to simulate geometrically intricate semiconductor devices with realistic models of the mobility and the recombination rate~\cite{Vasileska2010}. Its accuracy is demonstrated by comparing the results to those obtained by the FVM and FEM solvers implemented within the commercial software package COMSOL~\cite{COMSOL}. We should note here  that other DG schemes, such as discontinuous Petrov Galerkin~\cite{Causin2005}, hybridizable DG~\cite{Cockburn2009}, exponential fitted DG~\cite{Lombardi2012}, and DG with Lagrange multipliers~\cite{Borkera2017} could be adopted for the DG-based framework proposed in this work. 

The rest of the paper is organized as follows. Section II starts with the mathematical model where the coupled nonlinear system of Poisson and DD equations is introduced, then it describes the Gummel method and provides the details of the DG-based discretization. Section III demonstrates the accuracy and the applicability of the proposed framework via simulations of two realistic device examples. Finally, Section IV provides a summary and discusses possible future research directions.

\section{Formulation}
\label{sec:formulation}
\subsection{Mathematical Model}
The DD model describes the (semi-classical) transport of electrons and holes in an electric field under the drift-diffusion approximation~\cite{Selberher1984,Vasileska2010}. It couples the Poisson equation that  describes the behavior of the (static) electric potential and the two continuity equations that describe the behavior of electrons and holes. This (coupled) system of equations reads 
\begin{equation}
- \nabla  \cdot (\varepsilon ({\mathbf{r}})\nabla \varphi ({\mathbf{r}})) = q(C + {n_h}({\mathbf{r}}) - {n_e}({\mathbf{r}}))
\label{Poisson}
\end{equation}
\begin{equation}
\nabla  \cdot {{\mathbf{J}}_s}({\mathbf{r}}) =  \pm qR({n_e},{n_h}), s \in \{e,h\}
\label{DD0}
\end{equation}
where \({\mathbf{r}}\) represents the location vector, \({n_e}({\mathbf{r}})\) and \({n_h}({\mathbf{r}})\) are the electron and hole densities,  \( \varphi ({\mathbf{r}}) \) is the electric potential, \( {{\mathbf{J}}_e}({\mathbf{r}}) \) and \({{\mathbf{J}}_h}({\mathbf{r}})\) are the electron and hole current densities, \(\varepsilon ({\mathbf{r}})\) is the dielectric permittivity, $q$ is the electron charge, and \(R({n_e},{n_h})\) is the recombination rate. In \eqref{DD} and other equations in the rest of the text, $s \in \{e,h\}$, and the upper and lower signs should be selected for $s = e$  and $s = h$, respectively. The current densities \({{\mathbf{J}}_s}({\mathbf{r}})\) are given by
\begin{equation}
\,{{\mathbf{J}}_s}({\mathbf{r}}) = q{\mu _s}({\mathbf{E}}){\mathbf{E}}({\mathbf{r}}){n_s}({\mathbf{r}}) \pm q{d_s}({\mathbf{E}})\nabla {n_s}({\mathbf{r}}) 
\label{DDJ}
\end{equation}
where \({\mu _e}({\mathbf{E}})\) and \({\mu _h}({\mathbf{E}})\) are the (field-dependent) electron and hole mobilities, \({d_e}({\mathbf{E}}) = {V_T}{\mu _e}({\mathbf{E}})\)  and \({d_h}({\mathbf{E}}) = {V_T}{\mu _h}({\mathbf{E}})\) are the electron and hole diffusion coefficients, respectively, \({V_T} = {k_B}T/q\) is the thermal voltage, \({k_B}\) is the Boltzmann constant, $T$ is the absolute temperature, and 
\begin{equation}
{\mathbf{E}}({\mathbf{r}}) =  - \nabla \varphi ({\mathbf{r}})
\label{E}
\end{equation}
is the (static) electric field intensity. Inserting \eqref{DDJ} into \eqref{DD0} yields
\begin{align}
\nonumber \pm \nabla  \cdot ({\mu _s}({\mathbf{E}}){\mathbf{E}}({\mathbf{r}}){n_s}({\mathbf{r}})) + \nabla  \cdot ({d_s}({\mathbf{E}})\nabla {n_s}({\mathbf{r}})) \\
 \label{DD1} = R({n_e},{n_h}).
 \end{align}
The recombination rate \(R({n_e},{n_h})\) describes the recombination of carriers due to thermal excitation and various scattering effects. In this work, we consider the two most common processes, namely the trap assisted recombination described by the Shockley-Read-Hall (SRH) model~\cite{Vasileska2010} as
\begin{equation*}
{R_{\rm S\!R\!H}}({n_e},{n_h}) =\frac{ {n_e}({\mathbf{r}}){n_h}({\mathbf{r}}) - {n_i}^2 } { {\tau _e}({n_{h1}} + {n_h}({\mathbf{r}})) + {\tau _h}({n_{e1}} + {n_e}({\mathbf{r}})) }
\end{equation*}
and the three-particle band-to-band transition described by the Auger model~\cite{Vasileska2010} as
\begin{equation*}
{R_{\rm Auger}}({n_e},\!{n_h}) \! = \! ({n_e}({\mathbf{r}}){n_h}({\mathbf{r}}) - {n_i}^2)(C_e^A{n_e}({\mathbf{r}}) + C_h^A{n_h}({\mathbf{r}})).
\end{equation*}
Here, \({n_i}\) is the intrinsic carrier concentration, \({\tau _e}\) and \({\tau _h}\) are the carrier lifetimes, \({n_{e1}}\) and \({n_{h1}}\) are SRH model parameters related to the trap energy level, and \(C_e^A\) and \(C_h^A\) are the Auger coefficients. The net recombination rate \(R({n_e},{n_h})\) is given by~\cite{Vasileska2010}
\begin{equation}
R( {n_e}, {n_h}) = {R_{\rm S\!R\!H}}({n_e},{n_h}) + {R_{\rm Auger}}({n_e},{n_h})
\label{recombination}
\end{equation}
The mobility models have a significant impact on the accuracy of semiconductor device simulations. Various field- and temperature-dependent models have been developed for different semiconductor materials and different device operating conditions~\cite{Selberher1984,Vasileska2010,SILVACO,MINIMOS,COMSOL}. Often, high-field mobility models, which account for the carrier velocity saturation effect, are more accurate~\cite{Vasileska2010,SILVACO,MINIMOS,COMSOL}. In this work, we use the Caughey-Thomas model~\cite{Vasileska2010}, which expresses \({\mu _e}({\mathbf{E}})\) and \({\mu _h}({\mathbf{E}})\) as
\begin{equation}
{\mu _s}({\mathbf{E}}) = \mu _s^0{\left[ {1 + {{\left( {\frac{{\mu _s^0{E_\parallel }({\mathbf{r}})}}{{V_s^{sat}}}} \right)}^{{\beta _s}}}} \right]^{\beta _s^{ - 1}}}
\label{mobility}
\end{equation}
where \({E_\parallel }({\mathbf{r}})\) is amplitude of the electric field intensity parallel to the current flow, $\mu _e^0$ and $\mu _h^0$ are the low-field electron and hole mobilities, respectively, and \(V_s^{sat}\), \({\beta _e}\) and \({\beta _h}\) are fitting parameters obtained from experimental data.

\subsection{Gummel Method}
The DD model described by \eqref{Poisson}-\eqref{DD0} and \eqref{DDJ}-\eqref{E} represents a nonlinear and coupled system of equations. The electric field moves the carriers through the drift term in the expressions of \({{\mathbf{J}}_e}({\mathbf{r}})\) and \({{\mathbf{J}}_h}({\mathbf{r}})\)  [first term in \eqref{DDJ}]. The carrier movements change \({n_e}({\mathbf{r}})\) and \({n_h}({\mathbf{r}})\), which in turn affect \({\mathbf{E}}({\mathbf{r}})\) through the Poisson equation [see \eqref{Poisson}]. Furthermore, \(R({n_e},{n_h})\) [in \eqref{recombination}] and \({\mu _e}({\mathbf{E}})\) and \({\mu _h}({\mathbf{E}})\) [in \eqref{mobility}] are nonlinear functions of \({n_e}({\mathbf{r}})\) and \({n_h}({\mathbf{r}})\), and \({\mathbf{E}}({\mathbf{r}})\), respectively. This system can be solved using either a decoupled approach such as the Gummel method or a fully-coupled scheme such as the direct application of the Newton method~\cite{Vasileska2010,Farrell2018}. The Gummel method’s memory requirement and computational cost per iteration are less than those of the Newton method. In addition, accuracy and stability of the solution obtained by the Gummel method are less sensitive to the initial guess~\cite{Vasileska2010, Farrell2018}. On the other hand, the Gummel method converges slower, i.e., takes a higher number of iterations to converge to the solution~\cite{Vasileska2010, Farrell2018}. Since the simulations of the nanostructured devices considered in this work are memory-bounded, we prefer to use the Gummel method.

\begin{figure}[!t]
\centerline
{\includegraphics[width=0.5\columnwidth,draft=false]{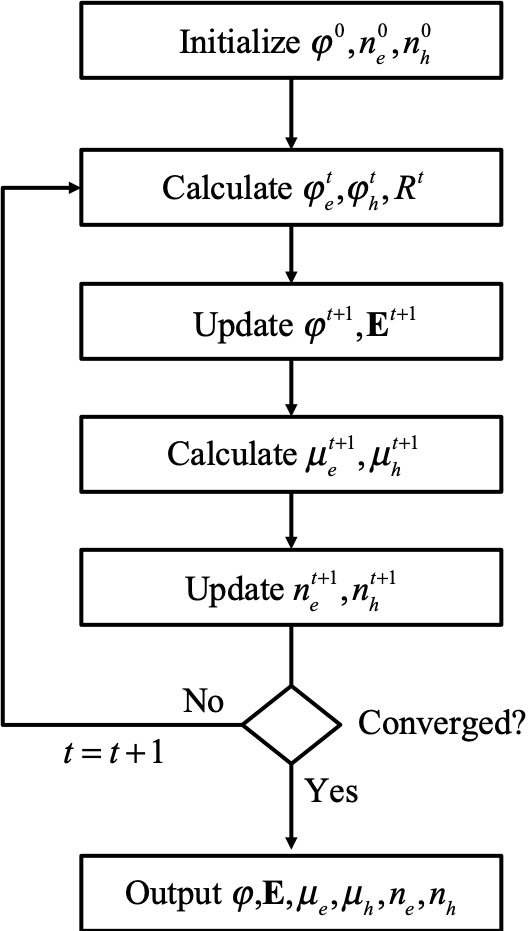}}
\caption{Gummel method.}
\label{Gummel}
\end{figure}

%\Figure[!t]()[width=0.95\textwidth]{imageName}
 %  {Caption.\label{fig:label}}

The Gummel iterations operate as described next and shown in Fig. \ref{Gummel}. To facilitate the algorithm, we first introduce the quasi-Fermi potentials~\cite{Selberher1984,Vasileska2010, Farrell2018}
\begin{equation}
{\varphi _s}({\mathbf{r}}) = \varphi ({\mathbf{r}}) \mp {V_T}\ln ({n_s}({\mathbf{r}})/{n_i}) , s \in \left\{ {e,h} \right\}.
\label{QuasiF}
\end{equation} 
``Inverting’’ \eqref{QuasiF} for \({n_e}({\mathbf{r}})\) and \({n_h}({\mathbf{r}})\), respectively, and inserting the resulting expressions into \eqref{Poisson} yield 
\begin{align}
\nonumber - \nabla \cdot (\varepsilon ({\mathbf{r}})\nabla \varphi ({\mathbf{r}})) = q(C & + {n_i}{e^{({\varphi _h}({\mathbf{r}}) - \varphi ({\mathbf{r}}))/{V_T}}}\\
\label{NLP} &  - {n_i}{e^{(\varphi ({\mathbf{r}}) - {\varphi _e}({\mathbf{r}}))/{V_T}}} ).
%\label{NLP} - \nabla \cdot (\varepsilon ({\mathbf{r}})\nabla \varphi ({\mathbf{r}})) &= q\left(C + {n_i}{e^{({\varphi _h}({\mathbf{r}}) - \varphi ({\mathbf{r}}))/{V_T}}}\right.\\
%\nonumber &  \left.- {n_i}{e^{(\varphi ({\mathbf{r}}) - {\varphi _e}({\mathbf{r}}))/{V_T}}}\right).
\end{align}
Equation \eqref{NLP} is termed as the nonlinear Poisson (NLP) equation simply because it is nonlinear in \(\varphi ({\mathbf{r}})\).  Using  \({\varphi _e}({\mathbf{r}})\) and \({\varphi _h}({\mathbf{r}})\), one can easily write the Frechet derivative of the NLP equation and solve the nonlinear problem with a fixed-point iteration technique such as the Newton method~\cite{Selberher1984, Vasileska2010, Farrell2018} (see below). The Gummel method decouples the NLP equation and the DD equations \eqref{DD0}; the nonlinearity is ``maintained’’ solely in the NLP equation and the DD equations are treated as linear systems~\cite{Selberher1984, Vasileska2010, Farrell2018} as shown by the description of the Gummel method below. 
To solve the NLP equation in \eqref{NLP}, we write it as a root-finding problem
\begin{align}
\nonumber & F( \varphi ({\mathbf{r}}), {\varphi _e}({\mathbf{r}}),{\varphi _h}({\mathbf{r}})) = \nabla  \cdot (\varepsilon ({\mathbf{r}})\nabla \varphi ({\mathbf{r}})) + \\
\label{F0} & q(C + {n_i}{e^{({\varphi _h}({\mathbf{r}}) - \varphi ({\mathbf{r}}))/{V_T}}} \! - \! {n_i}{e^{(\varphi ({\mathbf{r}}) - {\varphi _e}({\mathbf{r}}))/{V_T}}}) = 0.
\end{align}
The Frechet derivative of \(F(\varphi ({\mathbf{r}}),{\varphi _e}({\mathbf{r}}),{\varphi _h}({\mathbf{r}}))\) with respect to \(\varphi ({\mathbf{r}})\) is
\begin{align}
\nonumber F'(& \varphi ({\mathbf{r}}), {\varphi _e}({\mathbf{r}}), {\varphi _h}({\mathbf{r}});{\delta _\varphi }({\mathbf{r}})) = \nabla  \cdot (\varepsilon ({\mathbf{r}})\nabla {\delta _\varphi }({\mathbf{r}})) - \\
& {\textstyle{{q{n_i}} \over {{V_T}}}}({e^{({\varphi _h}({\mathbf{r}}) - \varphi ({\mathbf{r}}))/{V_T}}} + {e^{(\varphi ({\mathbf{r}}) - {\varphi _e}({\mathbf{r}}))/{V_T}}}){\delta _\varphi }({\mathbf{r}}).
\end{align}
The root-finding problem \eqref{F0} is solved iteratively as
\begin{equation}
\varphi {^{t+1}({\mathbf{r}})} = \varphi {^t({\mathbf{r}})} + {\delta _\varphi ^{t+1}}{({\mathbf{r}})}
\label{iter}
\end{equation}
where subscript ``$t$’’ refers to the variables at iteration $t$. In (12), \({\delta _\varphi ^{t+1}}{({\mathbf{r}})}\) is obtained by solving
\begin{align}
\nonumber F'(\varphi ^t {({\mathbf{r}})},{\varphi _e ^t}{({\mathbf{r}})},{\varphi _h ^t}{({\mathbf{r}})};  {\delta _\varphi ^{t+1}}{({\mathbf{r}})}) = \\
\label{F1} - F(\varphi ^t {({\mathbf{r}})},{\varphi _e ^t}{({\mathbf{r}})},{\varphi _h ^t}{({\mathbf{r}})})
\end{align}
where \(\varphi ^t {({\mathbf{r}})}\) is the solution at iteration $t$ (previous iteration), \({\varphi _e ^t}{({\mathbf{r}})}\) and \({\varphi _h ^t}{({\mathbf{r}})}\) are computed using using \(n_e^t({\mathbf{r}})\) and \(n_h^t({\mathbf{r}})\) in \eqref{QuasiF}. At iteration \(t=0\), initial guesses for  \(\varphi ^t {({\mathbf{r}})}\), \(n_e^t({\mathbf{r}})\) and \(n_h^t({\mathbf{r}})\) are used to start the iterations. Note that, in practice, one can directly compute  \(\varphi ^{t+1} {({\mathbf{r}})}\) without using the variable \({\delta _\varphi ^{t+1}}{({\mathbf{r}})}\). This is done by adding \(F'(\varphi ^t {({\mathbf{r}})},{\varphi _e ^t}{({\mathbf{r}})},{\varphi _h ^t}{({\mathbf{r}})};\varphi ^t {({\mathbf{r}})})\) to both sides of \eqref{F1}, and using \eqref{E} and the fact that
%\begin{equation*}
%\begin{aligned}
\begin{align*}
F'(\varphi ^t {({\mathbf{r}})},{\varphi _e ^t}{({\mathbf{r}})},{\varphi _h ^t} {({\mathbf{r}})};\varphi ^t {({\mathbf{r}})} + {\delta _\varphi ^{t+1}}{({\mathbf{r}})}) \\
 = F'(\varphi ^t {({\mathbf{r}})},{\varphi _e ^t}{({\mathbf{r}})},{\varphi _h ^t}{({\mathbf{r}})};\varphi ^{t+1} {({\mathbf{r}})})
\end{align*}
%\end{aligned}
%\end{equation*}
which result in the coupled system of equations in unknowns \({\phi ^{t+1}}({\mathbf{r}})\) and \({{\mathbf{E}}^{t+1}}({\mathbf{r}})\)
\begin{subequations} \label{NLP1}
\begin{align}
& \nabla  \cdot (\varepsilon ({\mathbf{r}}){{\mathbf{E}}^{t+1}}({\mathbf{r}})) + g({\mathbf{r}})\varphi ^{t+1} {({\mathbf{r}})} = f({\mathbf{r}}) \\
& {{\mathbf{E}}^{t+1}}({\mathbf{r}}) =  - \nabla {\varphi ^{t+1}}({\mathbf{r}}).
\end{align}
\end{subequations}
Here,
\begin{equation*}
g({\mathbf{r}}) = {\textstyle{{q{n_i}} \over {{V_T}}}}({e^{({\varphi _h ^t}{{({\mathbf{r}})}} - \varphi ^t {{({\mathbf{r}})}})/{V_T}}} + {e^{(\varphi ^t {{({\mathbf{r}})}} - {\varphi _e ^t}{{({\mathbf{r}})}})/{V_T}}})
\end{equation*}
and
\begin{align*}
f({\mathbf{r}}) = {\textstyle{{q{n_i}} \over {{V_T}}}}({e^{({\varphi _h ^t}{{({\mathbf{r}})}} - \varphi ^t {{({\mathbf{r}})}})/{V_T}}} + {e^{(\varphi ^t {{({\mathbf{r}})}} - {\varphi _e ^t}{{({\mathbf{r}})}})/{V_T}}}){\varphi ^k}({\mathbf{r}}) \\
+ q{n_i}(C/{n_i} + {e^{({\varphi _h ^t}{{({\mathbf{r}})}} - \varphi ^t {{({\mathbf{r}})}})/{V_T}}} - {e^{(\varphi ^t {{({\mathbf{r}})}} - {\varphi _e ^t}{{({\mathbf{r}})}})/{V_T}}})
\end{align*}
are known coefficients obtained from the previous iteration.

Unknowns  \(\varphi ^{t+1} {({\mathbf{r}})}\) and \(E^{t+1}{({\mathbf{r}})}\) are obtained by solving \eqref{NLP1}. Then, \({\mu _e}({{\mathbf{E}}^{t+1}})\) and \({\mu _h}({{\mathbf{E}}^{t+1}})\) are computed using \(E^{t+1}{({\mathbf{r}})}\) in \eqref{mobility}. Finally, \(n_e^{t+1}({\mathbf{r}})\) and \(n_h^{t+1}({\mathbf{r}})\) can be obtained by solving
\begin{align}
\nonumber  \pm \nabla  \cdot (&{\mu _s ^{t+1}}{({\mathbf{E}})} {\mathbf{E}}{^{t+1}({\mathbf{r}})}{n_s^{t+1}}{({\mathbf{r}})})  \\
\label{DD} + & \nabla  \cdot ({d_s^{t+1}}{({\mathbf{E}})}\nabla {n_s^{t+1}}{({\mathbf{r}})}) = R(n_e^t,n_h^t)
\end{align}
where \(R(n_e^t,n_h^t)\) on the right hand side is computed using \(n_e^t({\mathbf{r}})\) and \(n_h^t({\mathbf{r}})\) (from previous iteration) in \eqref{recombination}. Note that a ``lagging’’ technique may also be applied to \(R(n_e^t,n_h^t)\) to take advantage of the solutions at the current iteration. This technique expresses \(R({n_e},{n_h})\) as a summation of functions of \(n_e^t({\mathbf{r}})\) and \(n_h^t({\mathbf{r}})\) and \(n_e^{t+1}({\mathbf{r}})\) and \(n_h^{t+1}({\mathbf{r}})\), and moves the functions of \(n_e^{t+1}({\mathbf{r}})\) and \(n_h^{t+1}({\mathbf{r}})\) to the left hand side of \eqref{DD}. More details about this technique can be found in~\cite{Jerome1996}.

At this stage of the iteration, \({\varphi ^{t+1}}({\mathbf{r}})\), \(n_e^{t+1}({\mathbf{r}})\) and \(n_h^{t+1}({\mathbf{r}})\) are known; one can use these to compute \({\varphi _e ^{t+1}}{({\mathbf{r}})}\) and \({\varphi _h ^{t+1}}{({\mathbf{r}})}\) and move to the next iteration. Convergence of the iterations can be checked by either the residuals of \eqref{F0} and \eqref{DD} or by the difference between the solutions of two successive iterations.

\subsection{DG Discretization}
As explained in the previous section, at every iteration of the Gummel method, one needs to solve three linear systems of equations, namely \eqref{NLP1} and \eqref{DD} $(s=e,h)$. This can only be done numerically for arbitrarily shaped devices. To this end, we use the LDG method~\cite{Cockburn1998,Castillo2001} to discretize and numerically solve these equations. We start with the description of the discretization of \eqref{NLP1}.  First, we re-write \eqref{NLP1} in the form of the following boundary value problem
\begin{align}
\label{BVP0}
& \nabla  \cdot [\varepsilon ({\mathbf{r}}){\mathbf{E}}({\mathbf{r}})] + g({\mathbf{r}})\varphi ({\mathbf{r}}) = f({\mathbf{r}}),\quad {\mathbf{r}} \in \Omega\\
\label{BVP1}
& {\mathbf{E}}({\mathbf{r}}) =  - \nabla \varphi ({\mathbf{r}}),\quad {\mathbf{r}} \in \Omega\\
\label{BVP2}
& \varphi ({\mathbf{r}}) = {f_D}({\mathbf{r}}),\quad {\mathbf{r}} \in \;\partial {\Omega _D}\\
\label{BVP3}
& {\mathbf{\hat n}}({\mathbf{r}}) \cdot \varepsilon ({\mathbf{r}}){\mathbf{E}}({\mathbf{r}}) = {f_N}({\mathbf{r}}),\quad {\mathbf{r}} \in \;\partial {\Omega _N}.
\end{align}
In \eqref{BVP0}-\eqref{BVP3}, \(\varphi ({\mathbf{r}})\) and \({\mathbf{E}}({\mathbf{r}})\) are the unknowns to be solved for and \( \Omega \) is the solution domain. Note that in LDG, \({\mathbf{E}}({\mathbf{r}})\) is introduced as an auxiliary variable to reduce the order of the spatial derivative in \eqref{BVP0}. Here it is also a ``natural’’ unknown to be solved for within the Gummel method. Dirichlet and Neumann boundary conditions are enforced on surfaces  \(\partial {\Omega _D}\) and \(\partial {\Omega _N}\), and \({f_D}({\mathbf{r}})\) and \({f_N}({\mathbf{r}})\) are the coefficients associated with these boundary conditions, respectively. In \eqref{BVP3}, \({\mathbf{\hat n}} ({\mathbf r})\) denotes the outward normal vector \(\partial {\Omega _N}\) . For the problems considered in this work, \(\partial {\Omega _D}\) represents the metal contact surfaces with \({f_D}({\mathbf{r}}) = {V_{contact}}({\mathbf{r}})\), where \({V_{contact}}({\mathbf{r}})\) is the potential impressed on the contacts. The homogeneous Neumann boundary condition, i.e., \({f_N}({\mathbf{r}}) = 0\), is used to truncate the simulation domain~\cite{Schroeder1994}.

To facilitate the numerical solution of the boundary value problem described by \eqref{BVP0}-\eqref{BVP3} (within the Gummel method), \( \Omega \) is discretized into \(k\) non-overlapping tetrahedrons. The (volumetric) support of each of these elements is represented by \({\Omega _k}\), \(k = 1, \ldots ,K\). Furthermore, let \(\partial {\Omega _k}\) denote the surface of \({\Omega _k}\) and \({\mathbf{\hat n}}({\mathbf{r}})\) denote the outward unit vector normal to \(\partial {\Omega _k}\). Testing equations \eqref{BVP0} and \eqref{BVP1} with the Lagrange polynomials \({\ell _i}({\mathbf{r}})\), \(i = 1, \ldots ,{N_p}\),  on element \(k\) and applying the divergence theorem to the resulting equation yield the following weak form
\begin{align}
\nonumber & \int_{{\Omega _k}} {g({\mathbf{r}}){\varphi _k}({\mathbf{r}}){\ell _i}({\mathbf{r}})dV}  - \int_{{\Omega _k}} {\varepsilon ({\mathbf{r}}){{\mathbf{E}}_k}({\mathbf{r}}) \cdot \nabla {\ell _i}({\mathbf{r}})dV}  + \\
\label{PS0weak} & \oint_{\partial {\Omega _k}} \!\!\! { {\mathbf{\hat n}}({\mathbf{r}}) \cdot {{[\varepsilon ({\mathbf{r}}){{\mathbf{E}}_k}({\mathbf{r}})]}^*}{\ell _i}({\mathbf{r}})dS}  = \! \int_{{\Omega _k}} \!\!\! {f({\mathbf{r}}){\ell _i}({\mathbf{r}})dV}
\end{align}
  \vspace{-0.3cm}
\begin{align}
\nonumber \int_{{\Omega _k}} {{E_{k}^{\nu}}({\mathbf{r}}){\ell _i}({\mathbf{r}})dV}  - & \int_{{\Omega _k}} {{\varphi _k}({\mathbf{r}})\frac{\partial }{{\partial \nu }}{\ell _i}({\mathbf{r}})dV}  + \\
\label{PS1weak} & \oint_{\partial {\Omega _k}} { {{{\hat n}}_\nu}({\mathbf{r}}) {\varphi _k}{{({\mathbf{r}})}^*}{\ell _i}({\mathbf{r}})dS}  = 0.
\end{align}
Here, \({N_p}=(p+1)(p+2)(p+3)/6\) is the number of interpolating nodes,  \( p \) is the order of the Lagrange polynomials and subscript \( \nu \in \{ x,y,z\} \) is used for identifying the components of the vectors in the Cartesian coordinate system. We note here \({\varphi _k}({\mathbf{r}})\) and \({{\mathbf{E}}_k}({\mathbf{r}})\) denote the local solutions on element \(k\) and the global solutions on $\Omega$ are the sum of these local solutions.

\({\varphi ^*}\) and \({(\varepsilon {\mathbf{E}})^*}\) are numerical fluxes ``connecting’’ element \(k\) to its neighboring elements. Here, the variables are defined on the interface between elements and the dependency on \({\mathbf{r}}\) is dropped for simplicity of notation/presentation. In LDG, the alternate flux, which is defined as~\cite{Cockburn1998}
\begin{equation*}
{\varphi ^*} = \left\{ \varphi  \right\} + 0.5\boldsymbol{\hat \beta}  \cdot {\mathbf {\hat n} } \left[\kern-0.15em\left[ \varphi  
 \right]\kern-0.15em\right]
 \end{equation*}
 \begin{equation*}
{\left( {\varepsilon {\mathbf{E}}} \right)^*} = \left\{ {\varepsilon {\mathbf{E}}} \right\} - 0.5\boldsymbol{\hat \beta} ({\mathbf {\hat n} } \cdot \left[\kern-0.15em\left[ \varepsilon {\mathbf{E}}  \right]\kern-0.15em\right])
\end{equation*}
is used in the interior of \( \Omega \). Here, averaging operators are defined as \( \left\{ a \right\} = 0.5({a^ + } + {a^ - })\) and  \( \left\{ {\mathbf{a}} \right\} = 0.5({{\mathbf{a}}^ + } + {{\mathbf{a}}^ - })\) and  ``jumps’’ are defined as \( \left[\kern-0.15em\left[ a  \right]\kern-0.15em\right] = {a^ - } - {a^ + }\) and $\left[\kern-0.15em\left[ {\mathbf{a}} 
 \right]\kern-0.15em\right] = {{\mathbf{a}}^ - } - {{\mathbf{a}}^ + } \), where superscripts “-” and “+” refer to variables defined in element \(k\) and in its neighboring element, respectively. The vector \(\boldsymbol{\hat \beta} \) determines the upwinding direction of \(\varphi \) and \((\varepsilon {\mathbf{E}})\). In LDG, it is essential to choose opposite directions for \(\varphi \) and \((\varepsilon {\mathbf{E}})\), while the precise direction of each variable is not important~\cite{Cockburn1998, Hesthaven2008, Shu2016}. In this work, we choose \(\boldsymbol{\hat \beta}  = {\mathbf{\hat n}}\) on each element surface. On boundaries of \( \Omega \), the numerical fluxes are choosen as \( {\varphi ^*} = {f_D} \) and \( {\left( {\varepsilon {\mathbf{E}}} \right)^*} = {(\varepsilon {\mathbf{E}})^ - } \) on \( \partial {\Omega _D} \) and \( {\varphi ^*} = {\varphi ^-} \) and \( {\left( {\varepsilon {\mathbf{E}}} \right)^*} = {f_N} \) on \( \partial {\Omega _N} \), respectively~\cite{Castillo2001}.

We expand \({\varphi _k}({\mathbf{r}})\) and \({E_{k}^{\nu}}({\mathbf{r}})\) with the same set of Lagrange polynomials \({\ell _i}({\mathbf{r}})\)
\begin{equation}
{\varphi _k}({\mathbf{r}}) \simeq \sum\limits_{i = 1}^{{N_p}} {\varphi ({{\mathbf{r}}_i}){\ell _i}({\mathbf{r}})}  = \sum\limits_{i = 1}^{{N_p}} {\varphi _k^i{\ell _i}({\mathbf{r}})}
\label{expPhi}
\end{equation}
\begin{equation}
{E_{k}^{\nu}}({\mathbf{r}}) \simeq \mathop \sum \limits_{i = 1}^{{N_p}} {E_{\nu}}({{\mathbf{r}}_i}){\ell _i}({\mathbf{r}}) = \mathop \sum \limits_{i = 1}^{{N_p}} E_{k}^{\nu,i}{\ell _i}({\mathbf{r}})
\label{expE}
\end{equation}
where \({{\mathbf{r}}_i}\), $i = 1, \ldots, N_p$, denote the location of interpolating nodes, and $\varphi _k^i$ and $E_{k}^{\nu,i}$, $\nu \in \{ x,y,z\} $, $k = 1, \ldots, K$, are the unknown coefficients to be solved for. 

Substituting \eqref{expPhi} and \eqref{expE} into \eqref{PS0weak} and \eqref{PS1weak} for $k = 1, \ldots, K$, we obtain a global matrix system
\begin{equation}
\left[ {\begin{array}{*{20}{c}}
{ {\bar M^g} }&{\bar D \bar \varepsilon }\\
{\bar G}&{\bar M}
\end{array}} \right]
\left[ \begin{array}{l}
{\bar \Phi} \\
{\bar E}
\end{array} \right]
= \left[ \begin{array}{l}
{{\bar B}^{ \varphi} }\\
{{\bar B}^{{\mathbf E}}}
\end{array} \right].
\label{LS0}
\end{equation}
Here, the global unknown vectors \( \bar \Phi  = {[ {\bar \Phi _1}, \ldots, {\bar \Phi _K}]^T}\) and \(\bar E = {[ {\bar E_{1}^{x}},{\bar E_{1}^{y}}, {\bar E_{1}^{z}},...,{\bar E_{K}^{x}}, {\bar E_{K}^{y}}, {\bar E_{K}^{z}} ]^T}\) are assembled from elemental vectors  \({\bar \Phi _k} = [\varphi _k^1, ..., \varphi _k^{{N_p}}]\) and \({\bar E_{k}^{\nu}} = [E_{k}^{\nu,1},...,E_{k}^{\nu,N_p}]\), \( \nu \in \left\{ {x,y,z} \right\} \). The dimension of \eqref{LS0} can be further reduced by substituting \({\bar E} = {{\bar M}^{-1}}({\bar B}^{\mathbf E} - \bar{G} \bar \Phi )\) (from the second row) into the first row, which results in
\begin{equation}
({{\bar {M}}^g} - {\bar{D}} \bar \varepsilon {{\bar M}^{-1}}{\bar{G}}) {\bar \Phi}  = { {\bar B}^{\varphi} } - {\bar{D}} \bar \varepsilon {{\bar M}^{-1}}{ {\bar{B}}^{\mathbf E} }.
\label{LS1}
\end{equation}
In \eqref{LS0} and \eqref{LS1}, \({{\bar M}^g}\) and \({\bar M}\) are mass matrices. \({{\bar{M}}^g}\) is a $K \times K$ block diagonal matrix, where each ${N_p} \times {N_p}$ block is defined as 
\begin{equation*}
{\bar{M}}_{kk}^g(i,j) = \int_{{\Omega _k}} {g({\mathbf{r}}){\ell _i}({\mathbf{r}}){\ell _j}({\mathbf{r}})} dV.
\end{equation*}
\({\bar M}\) is also a $K \times K$ block diagonal matrix, where each block is a \(3 \times 3\) block diagonal matrix with ${N_p} \times {N_p}$ identical blocks defined as 
\begin{equation*}
{{\bar M}_{kk}^{(m)}}(i,j) = \int_{{\Omega _k}} {{\ell _i}({\mathbf{r}}){\ell _j}({\mathbf{r}})} dV, m=1,2,3.
\end{equation*}
\(\bar \varepsilon \) is a diagonal matrix with entries $ (\bar \varepsilon_1, \ldots, \bar \varepsilon_K) $, where $ \bar \varepsilon _{k} = ( \bar \varepsilon _{k}^{x}, \bar \varepsilon _{k}^{y}, \bar \varepsilon _{k}^{z} )$, ${\bar \varepsilon _{k}^{\nu}}(i) = {\varepsilon _k}({{\mathbf{r}}_i})$, ${k=1, \ldots, K}$, $\nu \in \left\{ {x,y,z} \right\}$. We note that $\varepsilon ({\mathbf{r}})$ is assumed isotropic and constant in each element.

Matrices \({\bar{G}}\) and \({\bar{D}}\) represent the gradient and divergence operators, respectively. For LDG, one can show that \({\bar D} =  - {{\bar G}^T}\)~\cite{Castillo2001}. The gradient matrix \({\bar G}\) is a $K \times K$ block sparse matrix, where each block is of size \(3{N_p} \times {N_p}\) and has contribution from the volume integral term and the surface integral term in \eqref{PS1weak}.
The volume integral term only contributes to diagonal blocks as \({\bar G}_{kk}^{vol} = {\left[ {{\bar S}_k^x\;{\bar S}_k^y\;{\bar S}_k^z} \right]^T}\), where
\begin{equation*}
{\bar S}_k ^\nu(i,j) =  - \int_{{\Omega _k}} { \frac{d{\ell _i}({\mathbf{r}})}{d\nu } {\ell _j}({\mathbf{r}}) }  dV, \nu \in \left\{ {x,y,z} \right\}.
\label{Svol}
\end{equation*}
The surface integral term contributes to both the diagonal blocks \({{\bar G}_{kk}}\) and off-diagonal blocks \({{\bar G}_{kk'}}\), where $k'$ corresponds to the index of the elements connected to element \(k\). Let \(\partial {\Omega _{kk'}}\) be the interface connecting elements \(k\) and $k'$, and let \({\theta _k}(j)\) select the interface nodes from element \(k\),
\begin{equation*}
{\theta _k}(j) = \left\{ \begin{array}{l}
1,\quad {{\mathbf{r}}_j} \in {\Omega _k},{{\mathbf{r}}_j} \in \partial {\Omega _{kk'}}\\
0,\quad otherwise
\end{array} \right..
\end{equation*}
Then, the contributions from the surface integral term to the diagonal block and the off-diagonal blocks are \({\bar G}_{kk}^{surf} = {\left[ {{\bar L}_k^x\;{\bar L}_k^y\;{\bar L}_k^z} \right]^T}\; \) and \({\bar G}_{kk'}^{surf} = {\left[ {{\bar L}_{k'}^x\;{\bar L}_{k'}^y\;{\bar L}_{k'}^z} \right]^T}\;\), where
\begin{align*}
%\label{Lkk} 
{\bar L}_k ^\nu(i,j) = \frac{{1 \! + \! sign(\boldsymbol{\hat \beta} \! \cdot \! \mathbf{\hat n})}}{2}{\theta _k}(j)\oint_{\partial {\Omega _{kk'}}} \!\!\! { {{\hat n}_\nu} ({\mathbf r}) {\ell _i}({\mathbf{r}}){\ell _j}({\mathbf{r}})dS}
\end{align*}
and
\begin{align*}
{\bar L}_{k'} ^{\nu}(i,j) = \frac{{1 \! - \! sign(\boldsymbol{\hat \beta} \! \cdot \! \mathbf{\hat n})}}{2}{\theta _{k'}}(j)\oint_{\partial {\Omega _{kk'}}} \!\!\! { {{\hat n}_\nu} ({\mathbf r}) {\ell _i}({\mathbf{r}}){\ell _j}({\mathbf{r}})dS}
%\label{Lkk1}
\end{align*}
respectively, \( \nu \in \left\{ {x,y,z} \right\} \).
The right hand side terms in \eqref{LS0} and \eqref{LS1} are contributed from the force term and boundary conditions and are expressed as
\begin{equation*}
{\bar B}_k^{\varphi} (i) = \int_{{\Omega _k}} {f({\mathbf{r}}){\ell _i}({\mathbf{r}})dV}  + \oint_{\partial {\Omega _k} \cap \partial {\Omega _N}} {{f_N}({\mathbf{r}}){\ell _i}({\mathbf{r}})dS}
\end{equation*}
\begin{equation*}
{\bar B}_{k}^{{\mathbf E},\nu}(i) = \oint_{\partial {\Omega _k} \cap \partial {\Omega _D}} {{{\hat n}_\nu({\mathbf r})}{f_D}({\mathbf{r}}){\ell _i}({\mathbf{r}})dS}, \nu \in \left\{ {x,y,z} \right\}.
\end{equation*}

The DD equations in \eqref{DD} (within the Gummel method) are also discretized using the LDG scheme as described next. Note that, here, we only discuss the discretization of the electron DD equation ($s = e$) and that of the hole DD equation ($s = h$) only differs by the sign in front of the drift term and the values of physical parameters. To simplify the notation/presentation, we drop the subscript denoting the species (electron and hole). The electron DD equation in \eqref{DD} is expressed  as the following boundary value problem
\begin{align}
\label{BVPDD0} & \nabla  \cdot [d({\mathbf{r}}){\mathbf{q}}({\mathbf{r}})]{\rm{ + }}\nabla  \cdot [{\mathbf{v}}({\mathbf{r}})n({\mathbf{r}})] = R({\mathbf{r}}),\quad {\mathbf{r}} \in \Omega\\
\label{BVPDD1} & {\mathbf{q}}({\mathbf{r}}) = \nabla n({\mathbf{r}}),\quad {\mathbf{r}} \in \Omega\\
\label{BVPDD2} & n({\mathbf{r}}) = {f_D}({\mathbf{r}}),\quad {\mathbf{r}} \in \;\partial {\Omega _D}\\
\label{BVPDD3} & {\mathbf{\hat n}} ({\mathbf{r}}) \cdot [d({\mathbf{r}}){\mathbf{q}}({\mathbf{r}}) + {\mathbf{v}}({\mathbf{r}})n({\mathbf{r}})] = {f_R}({\mathbf{r}}),\quad {\mathbf{r}} \in \;\partial {\Omega _R}.
\end{align}
Here \(n({\mathbf{r}})\) and \({\mathbf{q}}({\mathbf{r}})\) are the unknowns to be solved for and \( \Omega \) is the solution domain. The auxiliary variable \({\mathbf{q}}({\mathbf{r}})\) is introduced to reduce the order of the spatial derivative. \(d({\mathbf{r}}) = d({\mathbf{E}})\) and \({\mathbf{v}}({\mathbf{r}}) = \mu ({\mathbf{E}}){\mathbf{E}}({\mathbf{r}})\) become known coefficients during the solution of \eqref{DD} within the Gummel method. Dirichlet and Robin boundary conditions are enforced on surfaces \(\partial {\Omega _D}\) and \(\partial {\Omega _R}\), and \({f_D}({\mathbf{r}})\) and \({f_R}({\mathbf{r}})\) are the coefficients associated with these boundary conditions, respectively. \({\mathbf{\hat n}}({\mathbf{r}})\) denotes the outward normal vector of the surface. For the problems considered in this work,   represents electrode/semiconductor interfaces and, based on local charge neutrality~\cite{Schroeder1994}, \({f_D}({\mathbf{r}}) = (C + \sqrt {{C^2} + 4{n_i}^2} )/2\) and \({f_D}({\mathbf{r}}) = n_i^2/n_e^s\) for electron and hole DD equations, respectively. The homogeneous Robin boundary condition, i.e., \({f_R}({\mathbf{r}}) = 0\), is used on semiconductor/insulator interfaces, indicating no carrier spills out those interfaces~\cite{Schroeder1994}.

Following the same procedure used in the discretization of \eqref{NLP1}, we discretize the domain into non-overlapping tetrahedrons and test equations \eqref{BVPDD0} and \eqref{BVPDD1} using Lagrange polynomials on element \(k\). Applying the divergence theorem yield the following weak form:
\begin{align}
\nonumber \!-\!\int_{{\Omega _k}} \!\!\! {d({\mathbf{r}}){{\mathbf{q}}_k}({\mathbf{r}}) \! \cdot \! \nabla {\ell _i}({\mathbf{r}})dV}  & \!+\! \oint_{\partial {\Omega _k}} \!\!\!\!{ {\mathbf{\hat n}}({\mathbf r}) \cdot {{[d({\mathbf{r}}){{\mathbf{q}}_k}({\mathbf{r}})]}^*}{\ell _i}({\mathbf{r}})dS}\\
\nonumber \!-\!\int_{{\Omega _k}} \!\!\!\!{{\mathbf{v}}({\mathbf{r}}){n_k}({\mathbf{r}}) \cdot \nabla {\ell _i}({\mathbf{r}})dV} & \!+\! \oint_{\partial {\Omega _k}}\!\!\!\! { {\mathbf{\hat n}}({\mathbf r}) \cdot {{[{\mathbf{v}}({\mathbf{r}}){n_k}({\mathbf{r}})]}^*}{\ell _i}({\mathbf{r}})dS} \\
\label{DDweak0} &\!=\! \int_{{\Omega _k}} {f({\mathbf{r}}){\ell _i}({\mathbf{r}})dV} 
\end{align}
  \vspace{-0.3cm}
\begin{align}
\nonumber \int_{{\Omega _k}} {{q_{k}^{\nu}}({\mathbf{r}}){\ell _i}({\mathbf{r}})dV}  & + \int_{{\Omega _k}} {{n_k}({\mathbf{r}})\frac{\partial }{{\partial \nu }}{\ell _i}({\mathbf{r}})dV} \\
\label{DDweak1} & - \oint_{\partial {\Omega _k}} {{{{\hat n}}_\nu {(\mathbf{r})} }{n_k^*}{ ({\mathbf{r}}) }{\ell _i}({\mathbf{r}})dS}  = 0
\end{align}
where \( n^* \), \( (d{\mathbf{q}})^* \), and \( ({\mathbf{v}} n)^* \) are numerical fluxes ``connecting’’ element \(k\) to its neighboring elements. Here, for the simplicity of notation, we have dropped the explicit dependency on \( \mathbf{r} \) on element surfaces. For the diffusion term, the LDG alternate flux is used for the primary variable \({n^*}\) and the auxiliary variable \({(d{\mathbf{q}})^*}\)\cite{Cockburn1998}
\begin{equation*}
{n^*} = \left\{ n \right\} + 0.5\boldsymbol{\hat \beta}  \cdot {\mathbf {\hat n} } \left[\kern-0.15em\left[ n 
 \right]\kern-0.15em\right]
\end{equation*}
\begin{equation*}
{\left( {d{\mathbf{q}}} \right)^*} = \left\{ {d{\mathbf{q}}} \right\} - 0.5\boldsymbol{\hat \beta} ( {\mathbf {\hat n} } \cdot \left[\kern-0.15em\left[ d{\mathbf{q}}  \right]\kern-0.15em\right]).
\end{equation*}
Here, averages and ``jumps’’, and the vector coefficient \(\boldsymbol{\hat \beta} \) are same as those defined before. For the drift term, the local Lax-Friedrichs flux is used to mimic the path of information propagation~\cite{Hesthaven2008}
\begin{equation*}
{\left( {{\mathbf{v}}n} \right)^*} = \left\{ {{\mathbf{v}}n} \right\} + \alpha {\mathbf{\hat n}} ({n^- } - {n^+ }),\;
\alpha  = \frac{ \max ( \lvert \mathbf{\hat n} \cdot {{\mathbf{v}}^ - } \rvert, \lvert \mathbf{\hat n} \cdot {{\mathbf{v}}^ + } \rvert ) }{2}.
\end{equation*}
On boundaries, the numerical fluxes are choosen as \({n^*} = {f_D}\), \({\left( {d{\mathbf{q}}} \right)^*} = {(d{\mathbf{q}})^ - }\) and \({({\mathbf{v}}n)^*} = {\mathbf{v^-}}{f_D}\) on \(\partial {\Omega _D}\) and  \({n^*} = {n^ - }\) and \({(d{\mathbf{q}})^*} + {\left( { \mathbf{v} n } \right)^*} = {f_R}\) on \(\partial {\Omega _R}\), respectively. We note \({\left( {d{\mathbf{q}}} \right)^*}\) and \({({\mathbf{v}}n)^*}\) are not assigned independently on \(\partial {\Omega _R}\).

Expanding \({n_k}({\mathbf{r}})\) and \({q_{k}^{\nu}}({\mathbf{r}})\) with Lagrange polynomials \({\ell _i}({\mathbf{r}})\)
\begin{equation}
{n_k}({\mathbf{r}}) \simeq \sum\limits_{i = 1}^{{N_p}} {n({{\mathbf{r}}_i}){\ell _i}({\mathbf{r}})}  = \sum\limits_{i = 1}^{{N_p}} {n_k^i{\ell _i}({\mathbf{r}})}
\label{expN}
\end{equation}
\begin{equation}
{q_{k}^{\nu}}({\mathbf{r}}) \simeq \mathop \sum \limits_{i = 1}^{{N_p}} {q_v}({{\mathbf{r}}_i}){\ell _i}({\mathbf{r}}) = \mathop \sum \limits_{i = 1}^{{N_p}} q_{k}^{\nu,i}{\ell _i}({\mathbf{r}})
\label{expQ}
\end{equation}
where ${\mathbf{r}}_i$, $i=1, \ldots, N_p$, denote the location of interpolating nodes, $n_k^i$ and $q_{k}^{\nu,i}$, $\nu \in \{ x,y,z\} $ , $k = 1, \ldots, K$ are the unknown coefficients to be solved for. Substituting \eqref{expN} and \eqref{expQ} into \eqref{DDweak0} and \eqref{DDweak1}, we obtain a global matrix system
\begin{equation}
\left[ {\begin{array}{*{20}{c}}
{\bar{C}}&{{\bar{D}}\bar d}\\
{ - {\bar G}}&{{\bar M}}
\end{array}} \right]\left[ \begin{array}{l}
{\bar N}\\
{{\bar Q}}
\end{array} \right] = \left[ \begin{array}{l}
{{\bar B}^{n}}\\
{{\bar B}^{{\mathbf q}}}
\end{array} \right].
\label{LSDD0}
\end{equation}
Here, the global unknown vectors \({\bar N} = {[{{\bar N}_1},...,{{\bar N}_K}]^T}\) and \({\bar Q} = {[{{\bar{Q}}_{1}^{x}},{{\bar{Q}}_{1}^{y}},{{\bar{Q}}_{1}^{z}},...,{{\bar{Q}}_{K}^{x}},{{\bar{Q}}_{K}^{y}},{{\bar{Q}}_{K}^{z}}]^T}\) are assembled from elemental vectors \({{\bar N}_k} = [n_k^1,...,n_k^{{N_p}}]\) and \({{\bar{Q}}_{k}^{\nu}} = [q_{k}^{\nu,1},...,q_{k}^{\nu,N_p}], \nu \in \{x,y,z\} \). Substituting \({\bar Q} = {{\bar M}^{ - 1}}({{\bar B}^{{\mathbf q}}} + {\bar G \bar N})\) into \eqref{LSDD0} yields
\begin{equation}
({\bar{C}} + {\bar{D}}\bar d{{\bar M}^{ - 1}}{\bar G}){\bar N} = {{\bar B}^{n}} - {\bar{D}}\bar d{{\bar M}^{ - 1}}{{\bar B}^{{\mathbf q}}}
\label{LSDD1}
\end{equation}

In \eqref{LSDD0} and \eqref{LSDD1}, the mass matrix \({\bar M}\), the gradient matrix \({\bar G}\) and the divergence matrix \({\bar{D}}\) are same as those defined before. \(\bar d \) is a diagonal matrix with entries $ (\bar d_1, \ldots, \bar d_K) $, where $ \bar d _{k} = ( \bar d _{k}^{x}, \bar d _{k}^{y}, \bar d _{k}^{z} )$, ${\bar d _{k}^{\nu}}(i) = {d _k}({{\mathbf{r}}_i})$, ${k=1, \ldots, K}$, $\nu \in \left\{ {x,y,z} \right\}$. 

The block sparse matrix \({\bar{C}}\) has contribution from the third term (the volume integral) and the fourth term (the surface integral) in \eqref{DDweak0}. Each block is of size \({N_p} \times {N_p}\). The volume integral term only contributes to diagonal blocks as \( {\bar{C}}_{kk}^{vol} = \sum\nolimits_\nu  {{\bar{C}}_k ^\nu} \), where
\begin{equation*}
{\bar{C}}_k ^\nu(i,j) =  - \int_{{\Omega _k}} {{v_\nu }({\mathbf{r}}) \frac{d{\ell _i}({\mathbf{r}})}{d\nu}  {\ell _j}({\mathbf{r}})} dV, \nu  \in \{ x,y,z\}.
\end{equation*}
The surface integral term contributes to both the diagonal and off-diagonal blocks as 
\begin{align*}
& {\bar{C}}_{kk}^{surf}(i,j) =\\
&{\theta _k}(j) \oint_{\partial {\Omega _{kk'}}} { (\frac{1}{2}\sum\nolimits_\nu  {{{\hat n}_\nu}({\mathbf r}) v_{\nu}({\mathbf r})}  + \alpha({\mathbf r})) {\ell _i}({\mathbf{r}}){\ell _j}({\mathbf{r}})dS}
\end{align*}
and
\begin{equation*}
\begin{aligned}
& {\bar{C}}_{kk'}^{surf}(i,j) =\\
& {\theta _{k'}}(j) \oint_{\partial {\Omega _{kk'}}} { (\frac{1}{2}\sum\nolimits_\nu  {{{\hat n}_\nu}({\mathbf r}) v_{\nu}({\mathbf r})}  - \alpha({\mathbf r})) {\ell _i}({\mathbf{r}}){\ell _j}({\mathbf{r}})dS}
\end{aligned}
\end{equation*}
respectively, where $\nu  \in \{ x,y,z\}$, and $k'$, \(\partial {\Omega _{kk'}}\), and \({\theta _k}(j)\) are defined the same as before.

The right hand side terms in \eqref{LSDD0} are contributed from the force term and boundary conditions and are expressed as
\begin{align*}
{\bar B}_k^{n}(i)= & \int_{{\Omega _k}} {R({\mathbf{r}}){\ell _i}({\mathbf{r}})dV}+
\oint_{\partial {\Omega _k} \cap \partial {\Omega _R}} {{f_R}({\mathbf{r}}){\ell _i}({\mathbf{r}})dS}+\\
&\oint_{\partial {\Omega _k} \cap \partial {\Omega _D}} {{\mathbf{\hat n}} {(\mathbf{r})} \cdot {\mathbf{v}}({\mathbf{r}}){f_D}({\mathbf{r}}){\ell _i}({\mathbf{r}})dS}
\end{align*}
  \vspace{-0.3cm}
\begin{align*}
{\bar B}_{k}^{{\mathbf q},\nu}(i) = \oint_{\partial {\Omega _k} \cap \partial {\Omega _D}} {{{\hat n}_\nu({\mathbf r})}{f_D}({\mathbf{r}}){\ell _i}({\mathbf{r}})dS}, \nu  \in \{ x,y,z\}.
\end{align*}

\subsection{Sparse Linear Solver}
The sparse linear systems \eqref{LS1} and \eqref{LSDD1}  are constructed and solved in MATLAB. For small systems, one can use a direct solver. For large systems, when the number of unknowns is larger than 1\,000\,000 (when using double precison on a computer with 128GB RAM), it is preferable to use sparse iterative solvers to reduce the memory requirement. During our numerical experiments, we have found that the generalized minimum residual (GMRES) (the MATLAB built-in function ``gmres'’) outperforms other iterative solvers in execution time. Incomplete lower-upper (ILU) factorization is used to obtain a preconditioner for the iterative solver. The drop tolerance of the ILU is critical to keep the balance between the memory requirement and the convergence speed of the preconditioned iterative solution. A smaller drop tolerance usually results in a better preconditioner, however, it also increases the amount of fill-in, which increases the memory requirement.  

We note here that one can reuse the preconditioner throughout the Gummel iterations. Because the matrix coefficients change gradually between successive iterations, we can store the preconditioner obtained in the first iteration ($t=0$) and reuse it as the preconditioner in the following few iterations. In practice, the preconditioner only needs to be updated when the convergence of the sparse iterative solver becomes slower than it is in the previous Gummel iteration. For the devices considered in this work, the number of Gummel iterations is typically less than 50 and we find the preconditioners of the initial matrices work well throughout these iterations.

\section{Numerical Examples}

In this section, we demonstrate the accuracy and the applicability of the proposed DG-based framework by numerical experiments as detailed in the next two sections. We have simulated two practical devices and compared the results to those obtained by the COMSOL semiconductor module~\cite{COMSOL}.

\subsection{Metal-Oxide Field Effect Transistor}

First, we simulate a metal-oxide semiconductor field-effect transistor (MOSFET). The device is illustrated in Fig. 2. The background is uniformly-doped p-type silicon and source and drain regions are uniformly-doped n-type silicon. The doping concentrations in p- and n-type regions are ${10^{17}}{\rm cm^{-3}}$ and ${10^{18}}{\rm cm^{-3}}$, respectively. The source and drain are ideal Ohmic contacts attached to n-type regions. The gate contact is separated from the semiconductor regions by a silicon-oxide insulator layer. The dimensions of the device and the different material regions are shown in Fig. 3. Material parameters at $300$K are taken from~\cite{Levinshtein1996}.
\begin{figure}[!t]
\centerline{\includegraphics[width=0.95\columnwidth,draft=false]{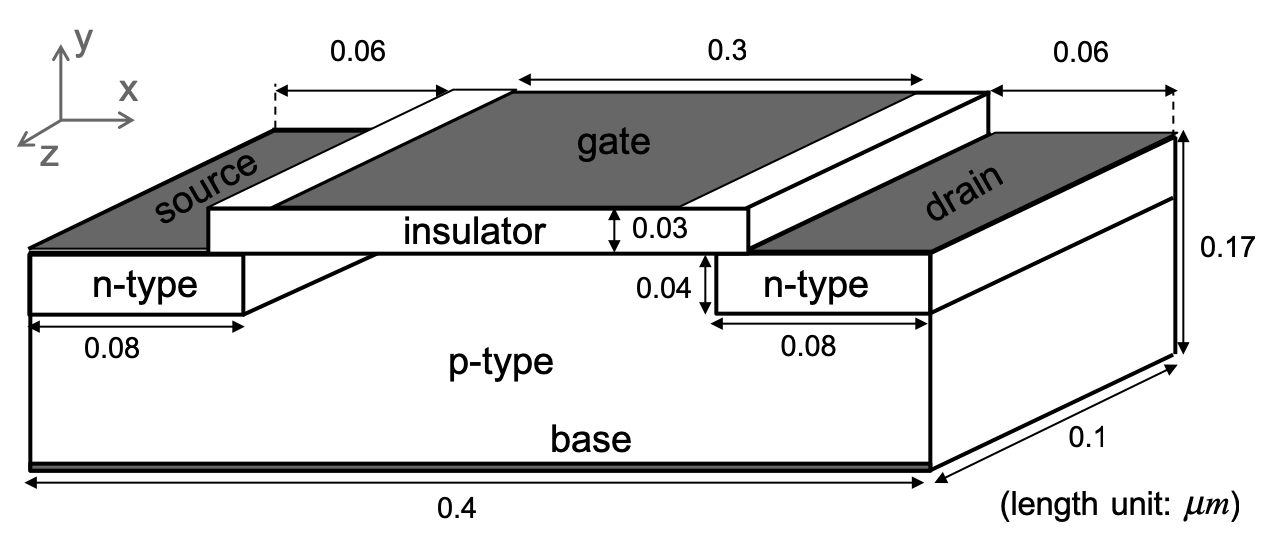}}
\caption{Schematic diagram of the MOSFET device.}
\label{MOSFET_Schem}
\end{figure}
\begin{figure}[!t]
  \centering
\subfloat[\label{3a}]{\includegraphics[width=0.95\columnwidth,draft=false]{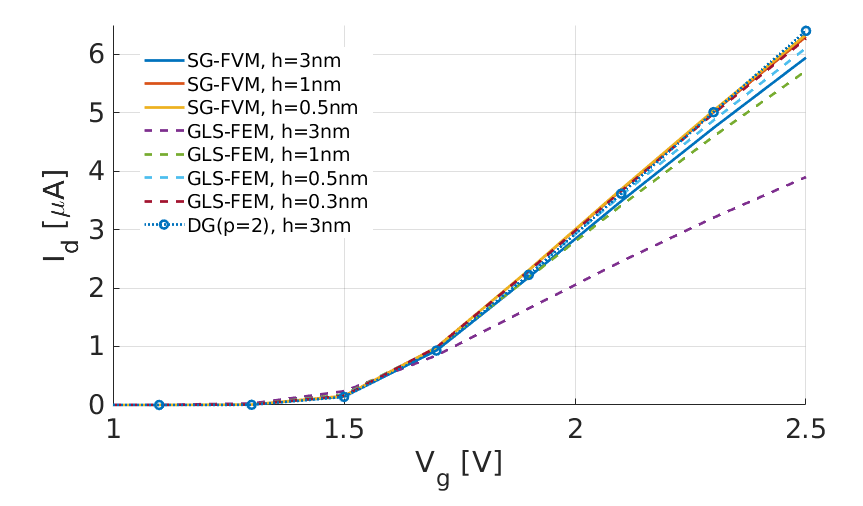}}\\
 \subfloat[\label{3b}]{\includegraphics[width=0.95\columnwidth,draft=false]{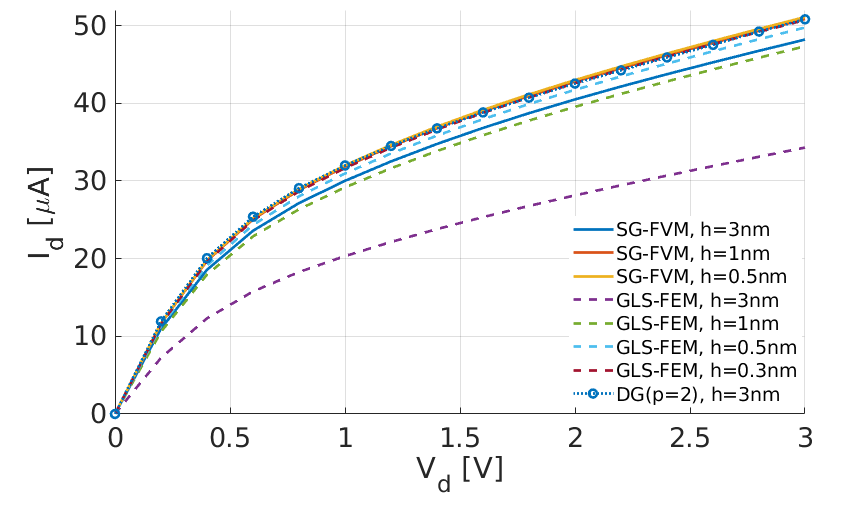}}
\caption{(a) The drain current $I_{d}$ versus gate voltage $V_{g}$ for drain voltage $V_{d}=0.1$V. (b) The drain current $I_{d}$ versus drain voltage $V_{d}$ for gate voltage $V_{g}=3$V.}
\label{MOSFET_IV}
\end{figure}

Special care needs to be taken to enforce the boundary conditions~\cite{Schroeder1994}. The DD equations are only solved in the semiconductor regions. Dirichlet boundary conditions are imposed on semiconductor-contact interfaces, where the electron and hole densities are determined from the local charge neutrality as ${n_e} = (C + \sqrt {{C^2} + 4{n_i}^2} )/2$ and ${n_h} = {n_i}^2/{n_e}$, respectively. Homogeneous Robin boundary condition, which enforces zero net current flow, is imposed on semiconductor-insulator interfaces. Poisson equation is solved in all regions. Dirichlet boundary conditions that enforce impressed external potentials are imposed on metal contacts (the contact barrier is ignored for simplicity). Homogeneous Neumann boundary condition is used on other exterior boundaries.

The semiconductor regions are discretized using a total of $122\,350$ elements and the order of basis functions in \eqref{expN} and \eqref{expQ} is 2. This makes the dimension of the system in \eqref{LSDD1} $1\,223\,500$. The regions where the Poisson equation is solved are discretized using a total of $170\,902$ elements and the order of basis functions in \eqref{expPhi} and \eqref{expE} is 2, making the dimension of the system in \eqref{LS1} $1\,709\,020$. The systems \eqref{LS1} and \eqref{LSDD1} are solved iteratively with a residual tolerance of $10^{-11}$. The drop tolerance of the ILU preconditioner is $10^{-5}$. The convergence tolerance of the Gummel method is $10^{-7}$.

Fig. 3 compares $I(V)$ curves computed using the proposed DG solver to those computed by the COMSOL semiconductor module. Note that this module includes two solvers: SG-FVM and GLS-FEM. For all three solvers, we refine the mesh until the corresponding $I(V)$ curve converges with a relative error of $10^{-2}$, where the error is defined as $\sum\nolimits_V {|I(V) - {I_{\rm ref}}(V)|} /\sum\nolimits_V {|{I_{\rm ref}}(V)|} $. Here, the reference ${I_{\rm ref}}(V)$ curve is obtained from the solution computed by the SG-FVM solver on a mesh with element size $h=0.5$nm. Fig. 3 shows that all $I(V)$ curves obtained by the three methods converge to $I_{\rm ref}(V)$ curve as the mesh they use is made finer. Fig. 3 (a) plots the drain current $I_{d}$ versus gate voltage $V_{g}$ under a constant drain voltage ${V_d} = 0.1$V. It shows that $I_{d}$ increases dramatically as $V_{g}$ becomes larger than a turn-on voltage ${V_{\rm th}}$ of approximately 1.5V. This indicates that a tunneling channel is formed between the source and the drain as expected. Fig. 3 (b) plots $I_{d}$ versus $V_{d}$ for $V_{g}=3$V.  It shows that $I_{d}$ increases continuously with $V_g$ and gradually saturates with a smaller slope of the $I(V)$ curve.

Comparing the $I(V)$ curves obtained by the three solvers using meshes with different element sizes, one can clearly see that the GLS-FEM solver requires considerably finer meshes than the SG-FVM and the DG solvers. The relative errors corresponding to different solvers and element sizes are listed in Table I. To reach a relative error of $10^{-2}$, the SG-FVM solver uses a mesh with $h=1$nm, the DG solver uses a mesh with $h=3$nm, while the GLS-FEM solver requires $h$ to be as small as $0.3$nm.
\begin{table}[!t]
\centering
\caption{Relative Error in $I(V)$ Curves}
\label{table}
\setlength{\tabcolsep}{3pt}
\begin{tabular}{|p{95pt}|p{55pt}|p{55pt}|}
\hline
& 
$I_d(V_g)$ & 
$I_d(V_d)$ \\ \hline
SG-FVM, $h=3$nm & 
5.83$\times 10^{-2}$& 
5.91$\times 10^{-2}$\\ \hline
SG-FVM, $h=1$nm & 
3.90$\times 10^{-3}$& 
4.97$\times 10^{-3}$ \\ \hline
GLS-FEM, $h=3$nm & 
3.35$\times 10^{-1}$& 
3.49$\times 10^{-1}$\\ \hline
GLS-FEM, $h=1$nm & 
7.83$\times 10^{-2}$& 
8.19$\times 10^{-2}$\\ \hline
GLS-FEM, $h=0.5$nm & 
2.94$\times 10^{-2}$& 
2.98$\times 10^{-2}$\\ \hline
GLS-FEM, $h=0.3$nm & 
9.11$\times 10^{-3}$& 
9.13$\times 10^{-3}$\\ \hline
DG, $h=3$nm & 
3.05$\times 10^{-3}$& 
5.28$\times 10^{-3}$\\ \hline
\end{tabular}
\label{tab1}
\end{table}

Figs. 4 (a) and (b) and Figs. 4 (c) and (d), respectively, compare the electron density and electric field intensity distributions computed by the DG and GLS-FEM solvers on the plane $z=0$ for $V_g=3$V and $V_d=0.5$V. Figs. 4(a) and (b) illustrate the “field-effect” introduced by the gate voltage, i.e., a sharp conducting channel forms near the top interface facing the gate ($y=0.2\mu {\rm m}$). This sharp boundary layer of carriers is the reason why a very fine mesh is required to obtain accurate results from this simulation. In Fig. 4 (b), the carrier density decays more slowly [compared to the result in Fig. 4(a)] away from the gate interface and suddenly drops to the Dirichlet boundary condition values at the bottom interface ($y=0$). This demonstrates the unphysical smearing of the boundary (carrier) layers observed in GLS-FEM solutions. Figs. 4 (c) and (d) show the $x$-component of the electric field intensity distribution computed by the DG and the GLS-FEM solvers, respectively. One can clearly see that the solution computed by the GLS-FEM solver is smoother (compared to the DG solution)  at the sharp corners of the gate.  The unphysical effects, as demonstrated in Figs. 4 (b) and (d), result from the GLS testing, which lacks of control on local conservation law~\cite{COMSOL}.
\begin{figure}[!t]
\centering
\subfloat[\label{4a}]{\includegraphics[width=0.76\columnwidth,draft=false]{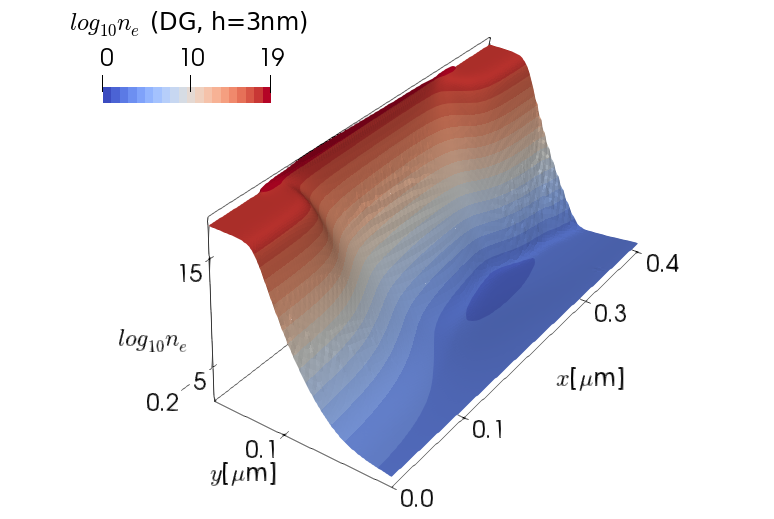}}\\
  \vspace{-0.3cm}
\subfloat[\label{4b}]{\includegraphics[width=0.76\columnwidth,draft=false]{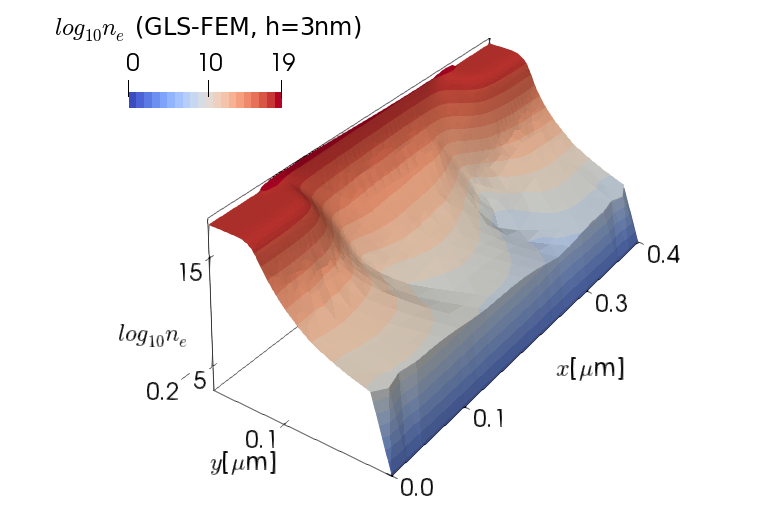}}\\
  \vspace{-0.3cm}
\subfloat[\label{4c}]{\includegraphics[width=0.75\columnwidth,draft=false]{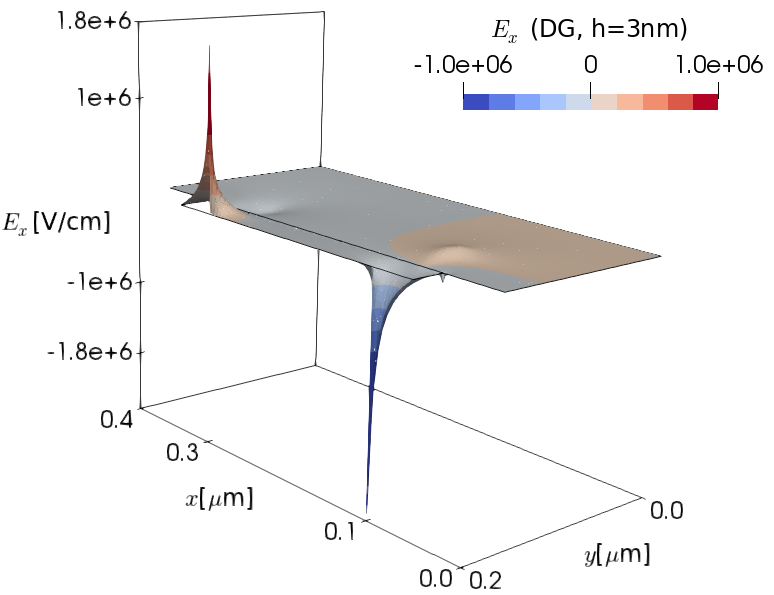}}\\
  \vspace{-0.3cm}
\subfloat[\label{4d}]{\includegraphics[width=0.75\columnwidth,draft=false]{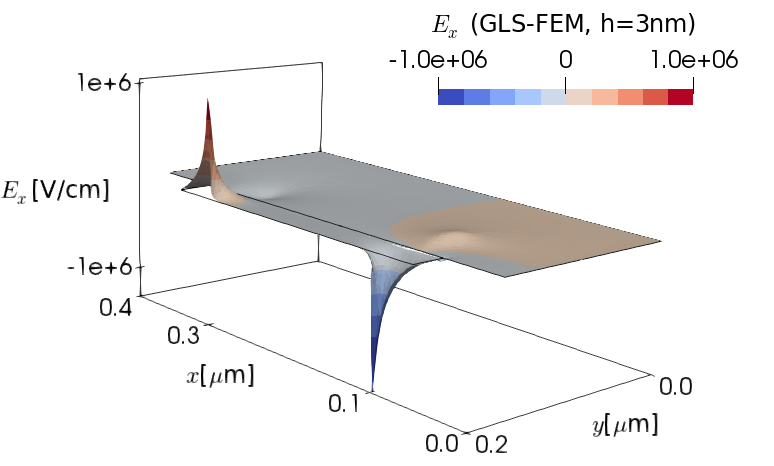}}
\caption{Electron density distribution computed on the plane $z=0$ by (a) the DG solver (b) the GLS-FEM solver for gate voltage $V_g=3$V and drain voltage $V_d=0.5$V. Electric field intensity distribution computed on the plane $z=0$ by (c) the DG solver (d) the GLS-FEM solver for gate voltage $V_g=3$V and drain voltage $V_d=0.5$V.}
\label{MOSFET_Ne_Ex}
\end{figure}

The SG-FVM solver requires the mesh to be admissible, which is often difficult to satisfy for 3D devices~\cite{Carey2000, Mauri2015, Farrell2018}. Implementation of SG-FVM in COMSOL uses a prism mesh generated by sweeping triangles defined on surfaces (for 3D devices)~\cite{COMSOL}. However, this leads to a considerable increase in the number of elements compared to the number of tetrahedral elements used by the DG and the SG-FEM solvers. In this example, the number of elements used by the SG-FVM is $545\,342$ ($h=1$nm), which results in $1\,499\,646$ unknowns. The DG solver refines the tetrahedral mesh near the boundaries where the solution changes fast, which is not possible to do using the prism mesh generated by sweeping triangles (Fig. 5). This mesh flexibility compensates for the larger number of unknowns required by the DG solver, which results from defining local expansions only connected by numerical flux.
\begin{figure}[ht]
  \centering  
  \subfloat[\label{5a}]{\includegraphics[width=0.8\columnwidth,draft=false]{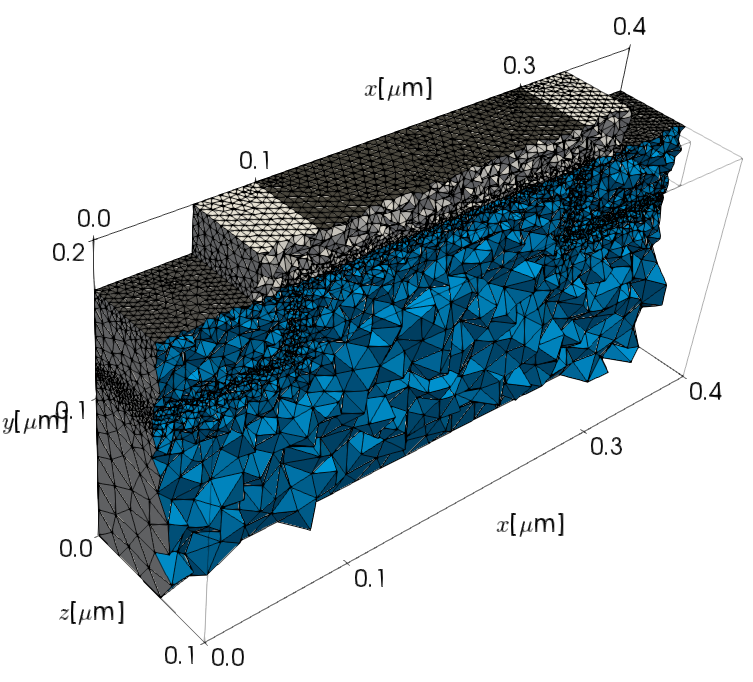}}\\
  \vspace{-0.3cm}
  \subfloat[\label{5b}]{\includegraphics[width=0.8\columnwidth,draft=false]{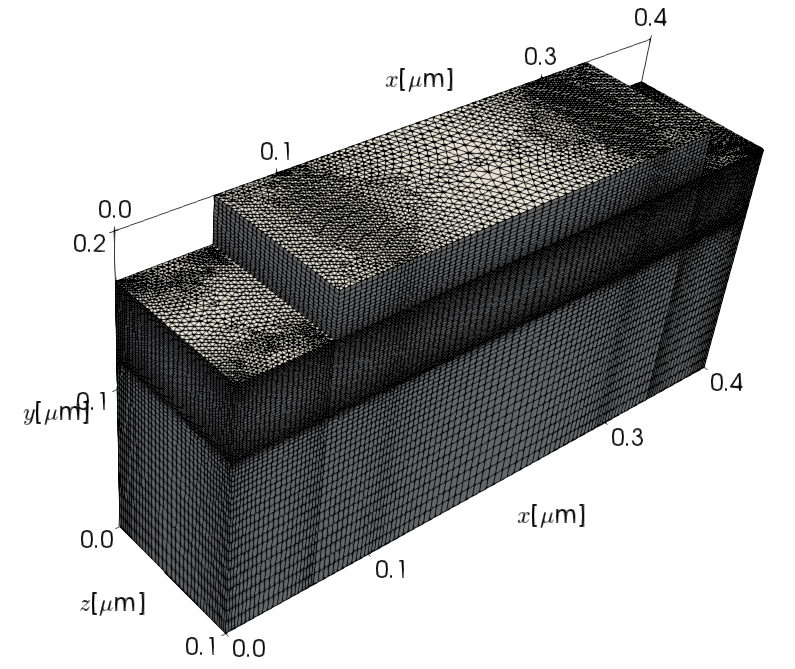}}
\caption{(a): Tetrahedral mesh used by DG and GLS-FEM. (b): Prism mesh used by SG-FVM in COMSOL\cite{COMSOL}.}
\label{MOSFET_Mesh}
\end{figure}

\subsection{Plasmonic-Enhanced Photoconductive Antenna}

For the second example, we consider a plasmonic-enhanced photoconductive antenna (PCA). The operation of a PCA relies on photoelectric effect: it ``absorbs’’ optical wave energy and generate  terahertz (THz) short-pulse currents. Plasmonic nanostructures dramatically enhances the optical-THz frequency conversion efficiency of the PCAs. The steady-state response of the PCAs, especially the static electric field and the mobility distribution in the device region, strongly influences their performance. Here, we use the proposed DG solver to simulate the device region of a PCA shown in Fig. 6, and compare the results to those obtained by the SG-FVM solver in COMSOL semiconductor module. 

Fig. 6 illustrates the device structure that is optimized to enhance the plasmonic fields near the operating optical frequency~\cite{Bashirpour2017}. The semiconductor layer is LT-GaAs that is uniformly doped with a concentration of ${10^{16}} {\rm cm}^{-3}$. The substrate layer is semi-insulating GaAs. We should note here that it is crucial to employ the appropriate field-dependent mobility models to accurately simulate this device~\cite{Moreno2014}. The Caughey-Thomas model is used here. Other material parameters same as those used in~\cite{Moreno2014}. The bias voltage is set to 10V.
\begin{figure}[!t]
\centerline{\includegraphics[width=0.95\columnwidth,draft=false]{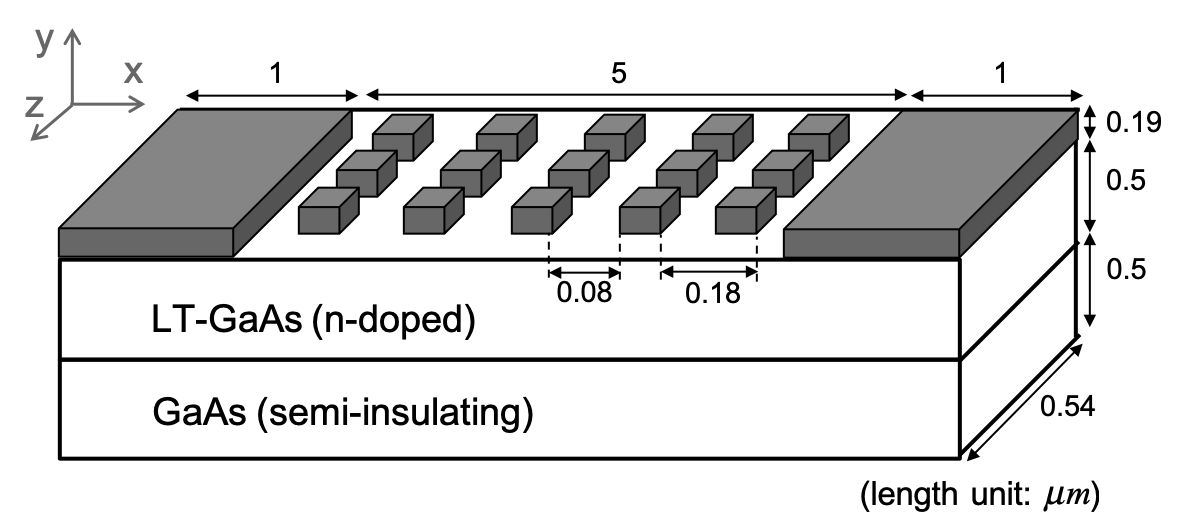}}
\caption{Schematic diagram of the plasmonic PCA.}
\label{PCA_Schem}
\end{figure}

The DD equations are solved in the semiconductor layer with Dirichlet boundary conditions on the semiconductor-contact interfaces and homogeneous Robin boundary condition on the semiconductor-insulator interfaces. Poisson equation is solved in the whole domain, which includes an extended air background. Dirichlet boundary conditions that enforce impressed external potentials are imposed on metal contacts. Floating potential condition is enforced on metals of the nanograting~\cite{Konrad1996}. Homogeneous Neumann boundary condition is used on exterior boundaries.

The semiconductor region is discretized using a total of $173\,711$ elements and the order of basis functions in \eqref{expN} and \eqref{expQ} is 2. This makes the dimension of the system in \eqref{LSDD1} $1\,737\,110$. The regions where the Poisson equation is solved are discretized using a total of $228\,921$ elements and the order of basis functions in \eqref{expPhi} and \eqref{expE} is 2, making the dimension of the system in \eqref{LS1} $2\,289\,210$. The systems \eqref{LS1} and \eqref{LSDD1} are solved iteratively with a residual tolerance of $10^{-11}$. The drop tolerance of the ILU preconditioner is $10^{-5}$. The convergence tolerance of the Gummel method is $10^{-7}$.

Fig. 7 (a) shows the electron density distribution computed by the proposed DG solver. Fig. 7 (b) plots the electron density computed by the DG and the SG-FVM solvers along lines $(x,y = 0.5\mu{\rm m}, z = 0)$ and $(x, y = 0,z = 0)$ versus $x$. The results agree well. The relative difference, which is defined as ${{{{\left\| {n_e^{\rm DG} - n_e^{\rm FVM}} \right\|}_2}} \mathord{\left/
 {\vphantom {{{{\left\| {n_e^{\rm DG} - n_e^{\rm FVM}} \right\|}_2}} {{{\left\| {n_e^{\rm FVM}} \right\|}_2}}}} \right.
 \kern-\nulldelimiterspace} {{{\left\| {n_e^{\rm FVM}} \right\|}_2}}}$, between the solutions obtained by the DG and the SG-FVM solvers is $0.78\%$. Here, ${\left\| {}. \right\|_2}$ denotes L2 norm and $n_e^{\rm DG}$ and $n_e^{\rm FVM}$ are the electron densities obtained by the two solvers. Note that $n_e^{\rm DG}$ is interpolated to the nodes where  $n_e^{\rm FVM}$ is computed.
\begin{figure}[!t]
  \centering
  \subfloat[\label{7a}]{\includegraphics[width=0.95\columnwidth,draft=false]{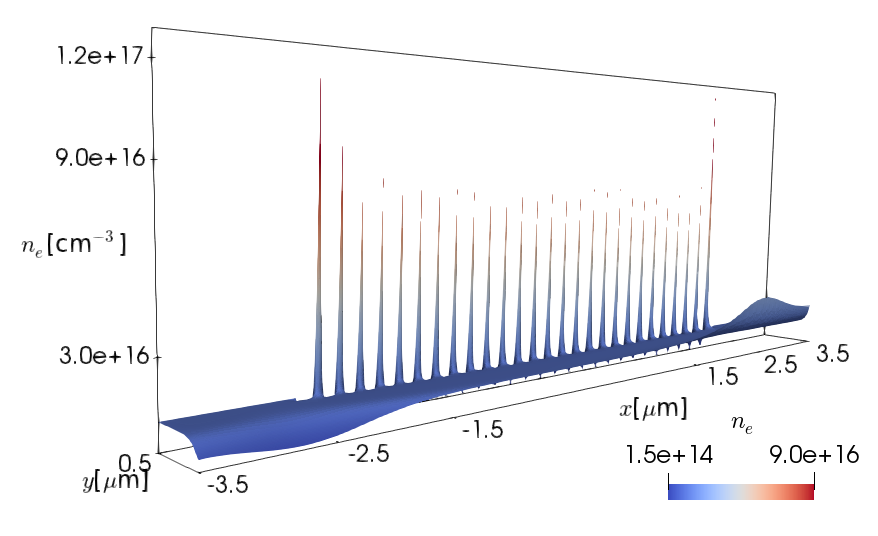}}\\
    \vspace{-0.3cm}
  \subfloat[\label{7b}]{\includegraphics[width=0.95\columnwidth,draft=false]{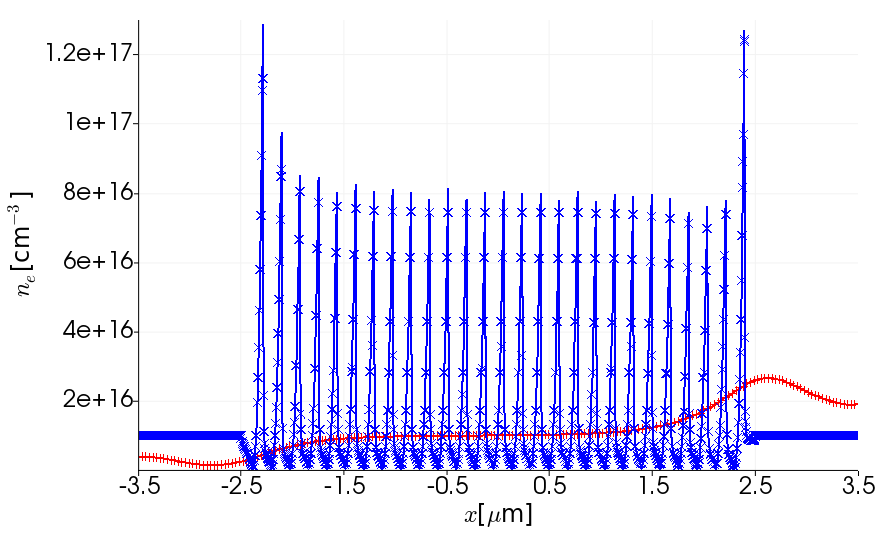}}
\caption{(a) Electron density distribution computed by the DG solver on the plane $z=0$ in the device region of the plasmonic PCA. (b) Electron density computed by the DG and the SG-FVM solvers along lines $(x,y = 0.5\mu{\rm m}, z = 0)$ and $(x, y = 0,z = 0)$ versus $x$.}
\label{PCA_Ne}
\end{figure}

\section{Conclusion}
In this paper, we report on a DG-based numerical framework for simulating steady-state response of geometrically intricate semiconductor devices with realistic models of mobility and recombination rate. The Gummel method is used to ``decouple’’ and ``linearize’’ the system of the Poisson equation (in electric potential) and the DD equations (in electron and hole charge densities). The resulting linear equations are discretized using the LDG scheme. The accuracy of this framework is validated by comparing simulation results to those obtained by the state-of-the-art FEM and FVM solvers implemented within the COMSOL semiconductor module. Just like FEM, the proposed DG solver is higher-order accurate but it does not require the stabilization techniques (such as GLS and SUPG), which are used by FEM. The main drawback of the proposed method is that it requires a larger number of unknowns than FEM for the same geometry mesh. But the difference in the number of unknowns gets smaller with the increasing order of basis function expansion. Additionally, DG can account for non-conformal meshes and benefit from local h-/p- refinement strategies. Indeed, we are currently working on a more ``flexible’’ version of the current DG scheme, which can account for multi-scale geometric features more efficiently by making use of these advantages. 

\section*{Acknowledgment}
The research reported in this publication is supported by the King Abdullah University of Science and Technology (KAUST) Office of Sponsored Research (OSR) under Award No 2016-CRG5-2953. Furthermore, the authors would like to thank the KAUST Supercomputing Laboratory (KSL) for providing the required computational resources.

\end{document}